\documentclass[twocolumn,aps,superscriptaddress,prl]{revtex4}
\usepackage{graphicx}
\usepackage{times}
\usepackage{amsmath}
\usepackage{amssymb}
\usepackage{units}
\usepackage{stmaryrd} 

\renewcommand{\vec}[1]{\boldsymbol{#1}}

\begin{document}
\preprint{0}

\title{A band structure scenario for the giant spin-orbit splitting observed at the Bi/Si(111) interface}

\author{Emmanouil Frantzeskakis}
\affiliation{Laboratoire de Spectroscopie Electronique, Institut
de Physique de la Mati\`{e}re Condens{\'e}e (ICMP), Ecole
Polytechnique F{\'e}d{\'e}rale de Lausanne (EPFL), station 3,
CH-1015 Lausanne, Switzerland}

\author{St{\'e}phane Pons}
\affiliation{Laboratoire de Spectroscopie Electronique, Institut
de Physique de la Mati\`{e}re Condens{\'e}e (ICMP), Ecole
Polytechnique F{\'e}d{\'e}rale de Lausanne (EPFL), station 3,
CH-1015 Lausanne, Switzerland}\affiliation{D{\'e}partement
Physique de la Mati\`{e}re et des Mat{\'e}riaux,
  Institut Jean Lamour, CNRS, Nancy Universit{\'e},
  F-54506 Vandoeuvre-les-Nancy, France}

\author{Marco Grioni}
\affiliation{Laboratoire de Spectroscopie Electronique, Institut
de Physique de la Mati\`{e}re Condens{\'e}e (ICMP), Ecole
Polytechnique F{\'e}d{\'e}rale de Lausanne (EPFL), station 3,
CH-1015 Lausanne, Switzerland}

\date{\today}

\begin{abstract}

\end{abstract}

\begin{abstract}
The Bi/Si(111) $(\sqrt{3}\times\sqrt{3})\textmd{R}30^{\circ}$ trimer phase offers a prime example of
a giant spin-orbit splitting of the electronic states at the interface with a semiconducting substrate.
We have performed a detailed angle-resolved photoemission (ARPES) study to clarify the complex topology of the hybrid interface bands.
The analysis of the ARPES data, guided by a model tight-binding
calculation, reveals a previously unexplored mechanism at the origin of the giant spin-orbit splitting, which relies
primarily on the underlying band structure. We anticipate that other similar interfaces characterized by trimer structures
could also exhibit a large effect.

\end{abstract}

\maketitle

\section{I. Introduction}

The normal spin degeneracy of the electronic states of
non-magnetic solids is lifted by the spin-orbit (SO) interaction
in crystals lacking an inversion center (Dresselhaus effect)
\cite{Dresselhaus1955,Konemann2005}. A similar effect was
predicted theoretically by Rashba and Bychkov (RB) for a
two-dimensional electron gas (2DEG) at a surface or an interface
which exhibits a structural surface asymmetry (SSA)
\cite{Bychkov1984}. Although the model was originally motivated by
semiconductor heterojunctions, split bands were first observed by
angle-resolved photoemission spectroscopy (ARPES) on metal
surfaces
\cite{LaShell1996,Rotenberg1999,Hochstrasser2002,Reinert2003,Koroteev2004,Henk2004,Malterre2007}.
Like its atomic counterpart, the RB effect has a relativistic
origin, namely the coupling of the spin to the magnetic field
which appears in the rest frame of the electron. In the
free-electron limit considered by RB, the parabolic dispersion is
split, as in Fig. 1, into two branches of opposite spin:
\begin{equation}
E^{\pm}(k)=\frac{\hbar^{2}k^{2}}{2m^{*}}\pm \alpha_R k \ ,
\end{equation}

where $k$ is the magnitude of the electron momentum in the plane
of the surface, and $m^*$ the effective mass. The Rashba parameter
$\alpha_\textmd{{R}}$ is proportional to the gradient of the
surface electric potential, and defines the strength of the RB
effect. The SO-splitting of the two branches can be quantified by
their momentum offset $2k_{0}=2\alpha_{R}m^{*}/\hbar^{2}$ or,
equivalently, by the Rashba energy
$E_{R}=\hbar^{2}k_{0}^{2}/(2m^{*})$, the difference between the
band minimum ($m^*>0$) or maximum ($m^*<0$) and the crossing point
of the two branches at $k=0$. In a more realistic approach, the
band splitting depends not only on the surface potential gradient,
but also on the atomic SO parameter, and on the asymmetry of the
electron wavefunctions
\cite{Petersen2000,Bihlmayer2006,Nagano2009,Malterre2007}.

\begin{figure}
\centering
  \includegraphics[width = 5.7 cm]{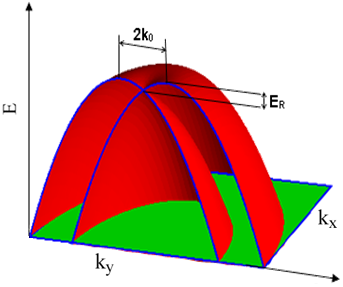}
  \caption{(color online) Schematics of the RB SO split bands for a 2D free electron gas}  \label{fig1}
\end{figure}

Interest in the RB effect has been revamped by observations of a
$giant$~SO-splitting in surface alloys formed by a high-Z element
-- Bi or Pb -- at the Ag(111) surface \cite{Ast2007,Meier2008}.
The unusual strength of the effect has prompted a reassessment of
the various factors contributing to the effect. It has been
suggested that additional components of the surface potential
gradient $within$ the surface, reflecting the anisotropic charge
distribution, are probably important \cite{Premper2007}.
Independent studies have stressed structural aspects, namely the
role of relaxation and buckling of the topmost layer in defining
the hybrid states \cite{Bihlmayer2007}.

Metallic surface alloys with a giant SO splitting are potentially
interesting for spintronics applications. The present challenge is
to make them compatible with semiconductor technology
\cite{Datta1990,Koo2009}. There has been encouraging progress in
this direction, and several attempts have been made to grow, on
Si(111) substrates, thin layers which support SO-split bands at
their surface
\cite{Frantzeskakis2008,Hirahara2006,He2008,FrantzeskakisJElSpectr2010,Dil2008}.
Recently, a giant spin-splitting with no buffer layers was
demonstrated for the isostructural Bi/Si(111)
\cite{Gierz2009,Sakamoto2009} and Bi/Ge(111) \cite{Hatta2009}
interfaces. In the same line, metallic spin-split surface states
were observed in the related Pb/Ge(111) system \cite{Yaji2010}. In
all cases, the electronic structure of these interfaces is more
complex than that predicted by the simple RB model. Although
first-principles calculations reproduce the experimental results,
they suffer from a certain lack of transparency. This leaves room
for a simpler but more direct approach which can help in the
interpretation of the experimental data, and thus contribute to
clarify the unconventional properties of the electronic states.
This was the motivation of the present work, which compares the
results of a detailed experimental band mapping of the Bi/Si(111)
interface by ARPES, with simple models of the band structure in
the presence of a RB-like interaction. In particular, a parametric
tight-binding scheme provides a satisfactory qualitative
description of the data, and suggests a possible new mechanism to
achieve a large spin polarization, which is closely connected with
a characteristic feature of the band structure of the interface.

\begin{figure}
  \centering
  \includegraphics[width = 8.7 cm]{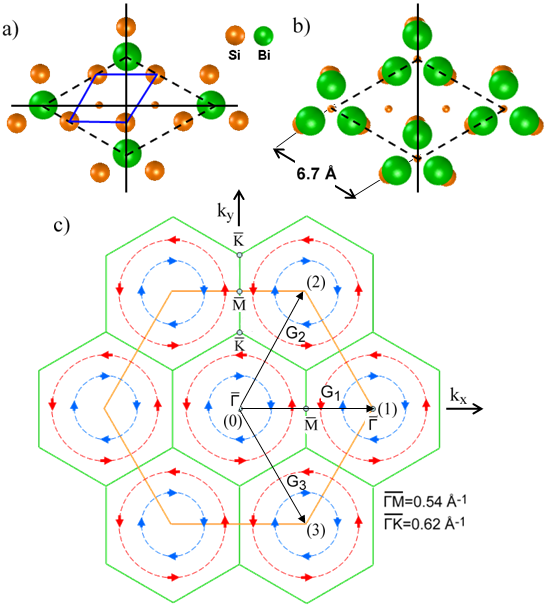}
  \caption{(color online)
The structure of (a) the monomer ($\alpha-$) and (b) the trimer
($\beta-\sqrt 3 \times \sqrt 3) \textmd{R} 30^{\circ}$ Bi/Si(111)
phases. The size of the Si atoms indicates their distance from the
surface. Solid (dashed) lines indicate the $1\times1$
$(\sqrt{3}\times\sqrt{3})$ primitive unit cell. The horizontal and
vertical black lines follow the mirror planes of the overlayers.
(c) The $1\times1$ (large hexagon) and
$(\sqrt{3}\times\sqrt{3})\textmd{R}30^{\circ}$ (small hexagons)
SBZs. The circles are constant-energy contours for RB paraboloids
centered at each equivalent $\overline{\Gamma}$ point of the
latter, and the arrows indicate the spin polarization. The
reciprocal lattice vectors $\mathbf{G}_{1}$, $\mathbf{G}_{2}$ and
$\mathbf{G}_{3}=\mathbf{G}_{1}-\mathbf{G}_{2}$ are those
considered in the NFE model of subsections III (B) and (C)}
\label{fig2}
\end{figure}

\section{II. Experimental Details}

The Si(111) substrate (Sb-doped, resistivity
0.01$\Omega.\textmd{cm}$) was flashed at 1200$^{\circ}$C by direct
current injection, and then cooled slowly in order to obtain a
sharp low-energy electron diffraction (LEED) (7 $\times$ 7)
pattern. The $(\sqrt 3 \times \sqrt 3) \textmd{R} 30^{\circ}$
Bi/Si(111) interface was prepared by deposition of $1$ monolayer
(ML) of Bi from a calibrated EFM3 Omicron source on the the
substrate at RT followed by a mild annealing. ARPES spectra were
acquired at 70 K and $21.2$ eV photon energy, with a PHOIBOS 150
SPECS Analyzer equipped with a monochromatized GammaData VUV 5000
high brightness source. The Fermi level position was determined
from the Fermi edge of a polycrystalline Au sample.

\section{III. Results and Model Calculations}

\subsection{A. ARPES Measurements}

\begin{figure}
  \centering
  \includegraphics[width = 8 cm]{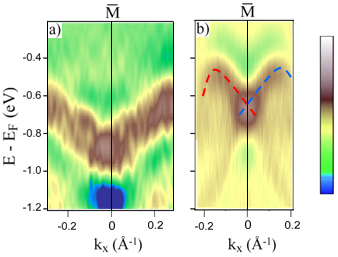}
  \caption{(color online):
2$^{\textmd{nd}}$
derivative ARPES intensity map along the
$\overline{\Gamma \textmd{M} \Gamma}$ direction for the monomer (a)
and the trimer (b) phases of Bi/Si(111).}   \label{fig3}
\end{figure}

The Bi/Si(111) interface exhibits two different structures with
the same  $(\sqrt{3}\times\sqrt{3})\textmd{R}30^{\circ}$
Bi/Si(111) periodicity: a monomer structure ($\alpha-$phase) for a
coverage of $1/3$ ML, and a trimer structure ($\beta-$phase) for
$1$ ML coverage. They are illustrated in Fig. 2 (a) and (b).
According to the widely accepted $\textmd{T}_{4}$ model
\cite{Nakatani1995}, both monomers and trimers are centered above
the 2$^{\textmd{nd}}$ layer Si substrate atoms. Figure 2(c) shows
the surface Brillouin zones (SBZ) of the unreconstructed Si(111)
surface, and of the $(\sqrt{3}\times\sqrt{3})\textmd{R}30^{\circ}$
superstructure. In the following we will always refer to the
latter. The $\alpha$ and $\beta$ phases have quite different band
structures, and can be easily distinguished. This is illustrated
in Fig. 3, where (second derivative) ARPES intensity maps of the
two phases are compared around the $\overline{\textmd{M}}$ point.
In agreement with previous studies, the $\alpha-$phase shows a
rather flat surface state with a broad minimum at
$\overline{\textmd{M}}$ \cite{Kim2002}, while the $\beta-$phase
shows two symmetrically dispersing features crossing at
$\overline{\textmd{M}}$ \cite{Kinoshita1987,Kim2001}. In the rest
of the paper we only consider the trimer $\beta-$phase, which is
the most interesting in the present context. The system has a
threefold rotation axis, and three mirror planes perpendicular to
the surface. One of these mirror planes is parallel to the
$[11\overline{2}]$ direction, and also to the
$\overline{\Gamma\textmd{KM}}$ direction of the SBZ ($k_{y}$ axis
in Fig. 2 (c)). The other mirror planes are rotated by
$120^{\circ}$ around the $z$ axis. The overlayer symmetry is
identical to the one of the substrate and corresponds to the plane
group $p31m$. The characteristic band crossing at
$\overline{\textmd{M}}$ is the signature of a peculiar RB-type
SO-splitting with a momentum offset $k_\textmd{{0}}$ of $0.126$
\AA$^{-1}$ and a Rashba energy of $140$ meV
\cite{Gierz2009,Sakamoto2009}. The large splitting has been
previously associated with the inversion asymmetry induced by the
trimers. We will see later that a somewhat different
interpretation is possible.

\begin{figure}
  \centering
  \includegraphics[width = 8.5 cm]{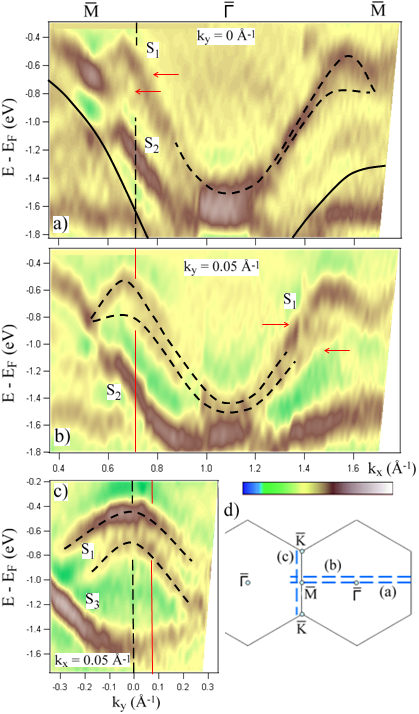}
  \caption{(color online)
(a) to (c) Second-derivative ARPES intensity maps showing the band
structure along the $k$-space cuts (a) to (c) of panel (d). The
dashed curves and arrows are guides to the eye, inspired by
results obtained in Refs. \cite{Gierz2009,Sakamoto2009}, and
highlight the dispersion of the S$_{1}$ split branches. The thick
solid line marks the edge of the projected Si bulk gap. Images (a)
and (b) intersect image (c) respectively along the vertical dashed
and solid lines. } \label{fig3}
\end{figure}

The ARPES intensity maps of Fig. 4 illustrate the dispersion of
three surface states -- labelled S$_{1}$ to S$_{3}$ as in Ref.
\onlinecite{Sakamoto2009} --. All three states are predicted to be
spin-polarized \cite{Sakamoto2009}. S$_{1}$ exhibits a large
splitting around $\overline{\textmd{M}}$, as already shown by Fig.
3, and a peculiar anisotropic dispersion around that point. The
experimental dispersion along the
$\overline{\Gamma\textmd{M}\Gamma}$ line (Fig. 4 (a)) shows a hint
of the two branches \cite{Gierz2009,Sakamoto2009}, which split
away from the crossing point at $\overline{\textmd{M}}$. The
dispersion is highlighted by dashed guides to the eye, which are
consistent with the results of Synchrotron Radiation (SR) studies,
where the individual S$_{1}$ components could be more clearly
resolved along $\overline{\Gamma\textmd{M}\Gamma}$
\cite{Sakamoto2009}. The two S$_{1}$ branches merge again
approaching the $\overline{\Gamma}$ point, where they cannot be
resolved from S$_{2}$. The energy splitting of the two branches
increases away from the $\overline{\Gamma\textmd{M}\Gamma}$
high-symmetry line, as shown by Fig. 4 (b) which shows the
dispersion along the parallel cut (b) of Fig. 4 (d) (i.e.
$k_\textmd{{y}}\neq0$). In support of our previous discussion on
the experimental results of Figs. 4 (a) and (b), the S$_{1}$ split
branches are easily identified in Fig. 4 (c), which shows a cut
parallel to the $\overline{\textmd{KMK}}$ line. Along this line
the intensity of S$_{2}$ is very small due to ARPES matrix
elements. The images of Fig. 4 (a) and (b) intersect the
perpendicular cut (c) along respectively the vertical black dashed
and the red solid lines.

\begin{figure}
  \centering
  \includegraphics[width = 9 cm]{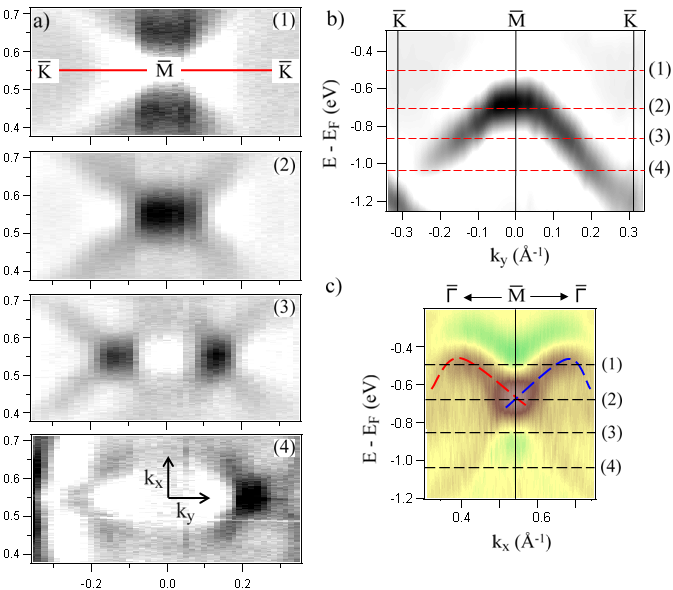}
  \caption{(color online)
(a) Experimental CE maps around $\overline{\textmd{M}}$. The
corresponding binding energies are $0.53$, $0.68$, $0.84$, and
$1.02$ eV for panels (1) to (4). They are indicated by horizontal
dashed lines in the intensity maps of panels (b) and (c), taken
along the $\overline{\textmd{KMK}}$ and
$\overline{\Gamma\textmd{M}\Gamma}$ directions. In the gray-scale
plots, highest intensity is black.} \label{fig5}
\end{figure}

Figure 5 illustrates the unusual topology of surface band S$_{1}$
around the $\overline{\textmd{M}}$ point, in a region of $k$-space
where it is well separated from other surface or bulk-derived
features. Panels 1 to 4 of Fig. 5 (a) show constant-energy (CE)
cuts taken at increasing binding energies between $0.53$ and
$1.02$ eV, corresponding to the horizontal dashed lines in the
intensity maps along the $\overline{\textmd{KMK}}$ and
$\overline{\Gamma\textmd{M}\Gamma}$ directions of panels (b) and
(c). Starting at the highest binding energy, the CE maps show two
intersections along $\overline{\textmd{KMK}}$, symmetrically
located with respect to $\overline{\textmd{M}}$. They get closer
at lower binding energy, following the negative-mass dispersion of
Fig. 5 (b). The two intersections finally merge at
$\overline{\textmd{M}}$ for E$_B\sim0.68$ eV. This is the maximum
of the dispersion along $\overline{\textmd{KMK}}$, and the
crossing point of the two SO-split branches along the
perpendicular $\overline{\Gamma\textmd{M}\Gamma}$ direction. Panel
1, taken above this energy, indeed shows non-intersecting CE
contours.

The data of Fig. 5 reveals that the topology of S$_1$ is quite
different from that of Fig. 1, predicted by the usual Rashba model
for a free-electron band centered at the $\overline{\Gamma}$
point. The degeneracy of the SO-split branches at
$\overline{\textmd{M}}$ is required by a combination of
time-reversal and translational symmetry. On the other hand, the
line of (near) spin degeneracy of Fig. 5 (b) finds no
correspondence in the simple Rashba model. The experimentally
observed large difference of the energy splitting along the two
high-symmetry directions is well captured by first-principles
calculations \cite{Gierz2009,Sakamoto2009}.

\subsection{B. An Isotropic Nearly-Free Electron Model}

We will now attempt a comparison of the experimental data using
simple theoretical models, to gain further insight in the unusual
dispersion of the SO-split bands. The minimal requirement for any
model is that it should include both the Rashba-like interaction
and translational invariance. The simplest approach satisfying
this condition is an isotropic nearly-free electron (NFE) model.
This model is schematically illustrated in Fig. 2 (c). RB
paraboloids ($m^*>0$) are centered at all equivalent
$\overline{\Gamma}$ points, and the CE lines are concentric
circles representing the inner and outer SO-split branches. The
spins exhibit a vortical structure around the $\overline{\Gamma}$
points. Two paraboloids centered at two adjacent SBZ intersect in
the common $\overline{\textmd{KMK}}$ Brillouin zone boundary along
two parabolas, one at lower energy for the outer SO branch, and a
second at higher energy for the inner.

The Rashba hamiltonian for a free-electron is \cite{Bychkov1984}:
\begin{equation}
H_{\textmd{RB}}(\vec{k})=
\alpha_R(\vec{\boldsymbol\sigma}\times\vec{k})_z\ \ \ ,
\end{equation}
where $\boldsymbol\sigma$ is the vector of Pauli matrices. In a
representation where the basis states are
$|\vec{k}\uparrow\rangle$, $|\vec{k}\downarrow\rangle$, and spin
projections refer to the $z$ axis, the corresponding matrix is:

\begin{equation}
H_{\textmd{RB}}(\vec{k})= \left(\begin{array}{cc}
           \hbar^{2}k^2/2m & \alpha_{R}(k_{y}+\imath k_{x}) \\
           \\
          \alpha_{R}(k_{y}-\imath k_{x}) & \hbar^{2}k^2/2m \\
\end{array}\right).
\end{equation}

Diagonalization of this hamiltonian generates SO-split paraboloids centered at $\overline{\Gamma}$ as in Fig. 1. The corresponding `spin-up' and `spin-down' eigenstates refer to a quantization axis $\mathbf{e}_{\theta}=\mathbf{e}_z\times\mathbf{k}/k$, which is always perpendicular to $\mathbf{k}$, i.e. tangential to the constant-energy circles. These states are therefore 100\% in-plane polarized, with a purely tangential spin polarization $\mathbf{P}$ -- opposite on the two branches -- rotating around $\overline{\Gamma}$.
This is easily generalized to include the lattice periodicity.
Since we are mainly interested in the band structure near E$_F$ around the $\overline{\textmd{M}}$ point, it is a good approximation,
to consider only the first SBZ and three adjacent zones centered at the  lattice vectors $\mathbf{G}_{1}=(1,0)$, $\mathbf{G}_{2}=(0,1)$ and
$\mathbf{G}_{3}=\mathbf{G}_{1}-\mathbf{G}_{2}$, as in Fig. 2 (c). The basis vectors, again referred to the $z$ axis,
are
$|\vec{k}\uparrow\rangle$, $|\vec{k}\downarrow\rangle$,
$|(\vec{k+G_{1}})\uparrow\rangle$,
$|(\vec{k+G_{1}})\downarrow\rangle$,
$|(\vec{k+G_{2}})\uparrow\rangle$,
$|(\vec{k+G_{2}})\downarrow\rangle$,
$|(\vec{k+G_{3}})\uparrow\rangle$,
$|(\vec{k+G_{3}})\downarrow\rangle$
The truncated NFE hamiltonian matrix is then:
\begin{widetext}
\begin{equation}
H_{\textmd{NFE}}= \\
\left(%
\begin{array}{cccc}
  H_{\textmd{RB}}(\vec{k}) & V_{01} & V_{02} & V_{03} \\
  V^{*}_{01} & H_{\textmd{RB}}(\vec{k+G_{1}}) & V_{12} & V_{13} \\
  V^{*}_{02} & V^{*}_{12} & H_{\textmd{RB}}(\vec{k+G_{2}}) & V_{23} \\
  V^{*}_{03} & V^{*}_{13} & V^{*}_{23} & H_{\textmd{RB}}(\vec{k+G_{3}}) \\
\end{array}%
\right).
\end{equation}
\end{widetext}

The diagonal $2\times2$ building blocks now generate SO-split
paraboloids centered at $\overline{\Gamma}$, and at
$\mathbf{G}_{1}$, $\mathbf{G}_{2}$ and $\mathbf{G}_{3}$. The
off-diagonal $2\times2$ blocks describe the interaction between
states of equal spin on the various paraboloids \cite{Didiot2006}.
The hybridization strength is, as usual,  the corresponding
Fourier component of the crystal potential $V(r)$ defined by:

\begin{equation}
V(r)=\sum_{i}V_{G_{i}}e^{\imath \vec{G}_{i}\cdot \vec{r}}\ \ \ .
\end{equation}
For instance:
\begin{equation}
V_{01}= \left(\begin{array}{cc}
           V_{G_{1}}& 0 \\
           0&V_{G_{1}} \\
\end{array}\right)\ \ \ .
\end{equation}
It is easy to show that (6) is equivalent to an interaction of the
form $V(\mathbf{k},\mathbf{k+G_{1}})=V_{G_1}$cos$(\delta/2)$,
where $\delta$ is the angle between the two polarization vectors.
In our case all hybridization terms are equal: $V_{Gi}=V_G$, and
$H_{NFE}$ contains only the two parameters $\alpha_R$ and $V_G$.
The binding energy of the paraboloids at $\overline{\Gamma}$, or
equivalently the Fermi level position, are adjusted to fit the
experimental data of Fig. 4.

\begin{figure}
  \centering
  \includegraphics[width = 8.5 cm]{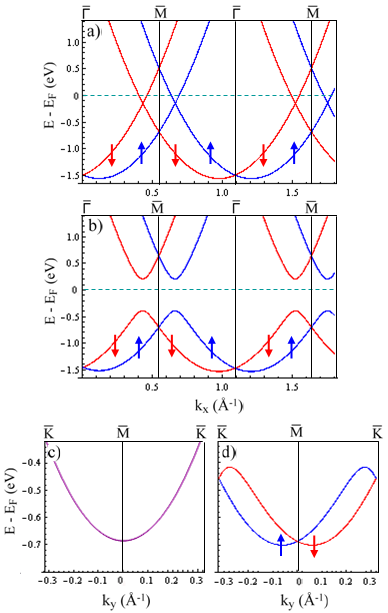}
  \caption{(color online) (a), (b) Band dispersion of the isotropic NFE model along the $\overline{\Gamma\textmd{M}\Gamma}$ for
  $|V_{G}|=0$ and $|V_{G}|=0.3$ eV. Arrows indicate the opposite (in-plane) spin polarization of the two branches.
  (c), (d) Same for the $\overline{\textmd{KMK}}$ high-symmetry direction. The two spin states are degenerate along the SBZ boundary for
  $|V_{G}|=0$.  Notice the different scales in (a), (b) and (c), (d).
  } \label{fig6}
\end{figure}

Figure 6 (a) illustrates the predictions of the NFE model along
the $\overline{\Gamma\textmd{M}\Gamma}$ direction, in the limit
$|V_{G}|=0$. The outer branches cross at the
$\overline{\textmd{M}}$ point, with opposite spin polarization.
Along the perpendicular $\overline{\textmd{KMK}}$ direction (Fig.
6 (c)), their intersection is a parabola dispersing upwards from
$\overline{\textmd{M}}$. The inner branches of the paraboloids
similarly cross above E$_{\textmd{F}}$. The model yields a
constant momentum separation between the SO-split branches of each
paraboloid, and does not capture the experimentally observed
$k$-dependent splitting. Moreover, the sign of the dispersion
along the SBZ boundary is opposite to that of Fig. 5 (b), and the
opposite spin states are strictly degenerate, for any value of
$\alpha_R$.

Figure 6 (b) shows that the main effect of a finite lattice
potential is the opening of energy gaps at the crossing of bands
with parallel spins, i.e. at the crossing of the outer branch of
one paraboloid with the inner branch of the paraboloid centered at
an adjacent SBZ. No gap opens when either the outer or the inner
branches cross at the $\overline{\textmd{M}}$ point, because their
spins are opposite there. This is consistent with the requirements
of time-reversal symmetry. At the other crossing points along the
SBZ boundary the spins are not strictly opposite, but they are
nonetheless rather antiparallel, and the hybridization is
therefore small. Therefore, as shown in Fig. 6 (d), the two spin
states are not degenerate as in the case $|V_{G}|=0$, but their
energy separation is small along this direction. The model again
yields a positive effective mass along $\overline{\textmd{KMK}}$,
in contrast with the experiment. The momentum splitting along
$\overline{\textmd{KMK}}$ scales with $|V_{G}|$ and
$\alpha^{1/2}$, confirming the unconventional character of the
underlying mechanism. It should also be noted that although the
$\overline{\textmd{M}}$ spin degeneracy is fundamental, the
degeneracy predicted at $\overline{\textmd{K}}$ is accidental. It
is lifted when further reciprocal lattice points are included.

\begin{figure}
  \centering
  \includegraphics[width = 8 cm]{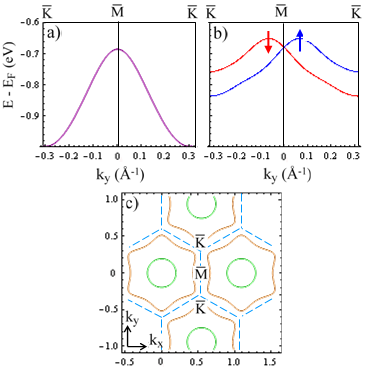}
  \caption{(color online) (a), (b) Band dispersion of the anisotropic NFE model along the $\overline{\textmd{KMK}}$
  SBZ boundary for $|V_{G}|=0$ and $|V_{G}|=0.3$ eV.
  (c) Constant energy contours for a binding energy of $1.1$ eV showing the anisotropic shape of the spin-split states. Green (light gray) and brown (dark gray) colors indicate positive and negative values of tangential spin polarization.} \label{fig7}
\end{figure}

\subsection{C. An Anisotropic Nearly-Free Electron Model}

The positive effective mass along the Bragg planes predicted in
Fig. 6 (c), (d)  is a direct consequence of the simple circular CE
contours of the isotropic NFE model. More complex contours are
however possible in anisotropic 2D systems. Blossom-like contours
have been predicted for Rashba systems \cite{Premper2007,Ast2007},
and experimentally observed for the BiAg$_{2}$ \cite{Gierz2009}
and SbAg$_{2}$ \cite{MoreschiniSb2009} surface alloys. Concave CE
contours are a generic effect of an in-plane anisotropy of the
potential and they are not limited to surface alloys. For example,
a 2D Dirac fermion state at the surface of the topological
insulator Bi$_{2}$Te$_{3}$ has been shown to exhibit a
snow-flake-like Fermi surface \cite{Fu2009}.

\begin{figure*}
  \centering
  \includegraphics[width = 18 cm]{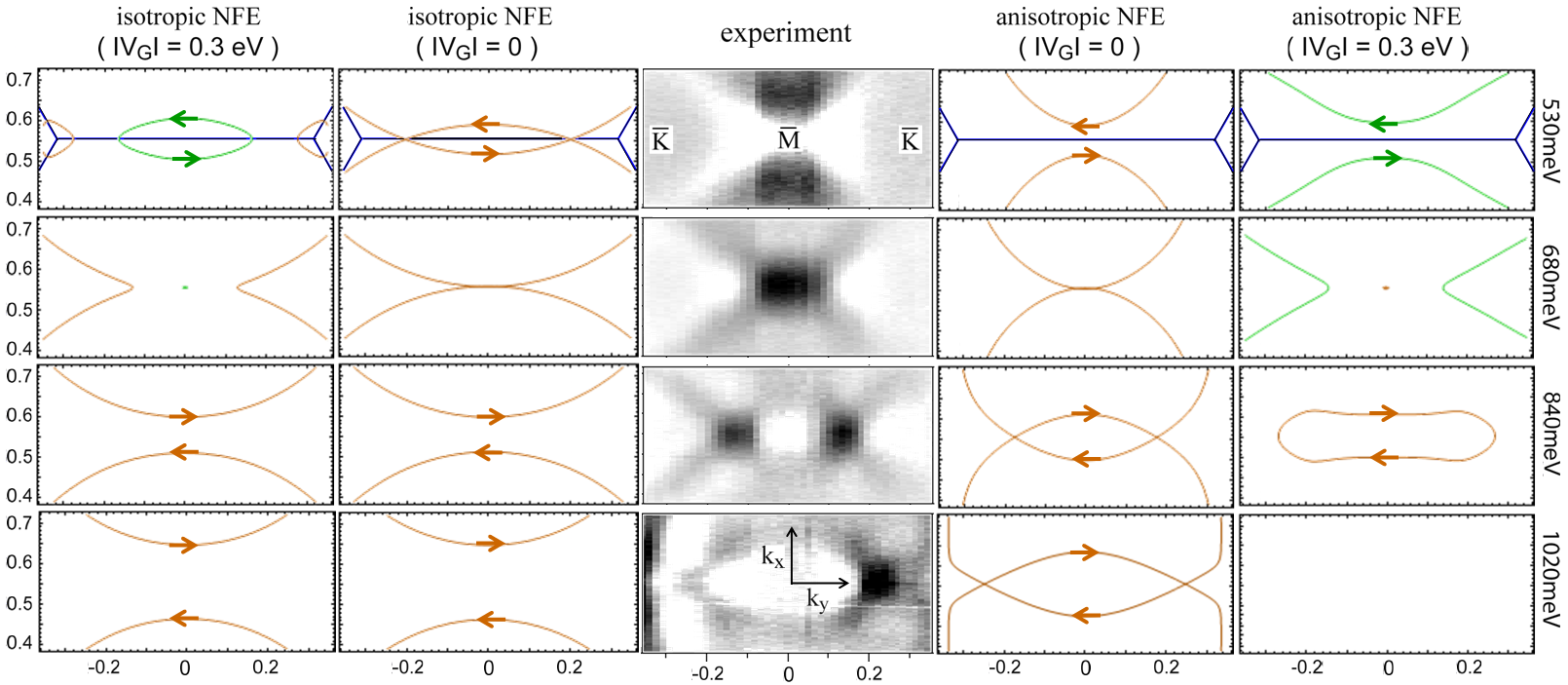}
  \caption{(color online) Constant energy maps as measured by ARPES (middle panel) and according to the predictions of the isotropic (left panels)
  and the anisotropic (right panels) NFE models described in the text. Both models are presented for $|V_{G}|=0$ and $|V_{G}|=0.3$ eV. Green (light gray) and brown (dark gray) colors indicate positive and negative values of tangential spin polarization. The arrows are a sketch of
  the predicted in-plane projection of the spin polarization around the $\overline{\textmd{M}}$ point.}
\label{fig8}
\end{figure*}

$\mathbf{k}\cdot \mathbf{p}$ theory has been used to calculate
higher-order terms in the effective Hamiltonian of a topological
insulator with $R3\bar{m}$ symmetry \cite{Fu2009}. Following these
results we introduce an anisotropy in $H_{\textmd{NFE}}(k)$ as:
\begin{equation}
H_{\textmd{an}}(k)=H_{\textmd{NFE}}(k)+H'(k)\ \ ,
\end{equation}
with
\begin{equation}
H'(k)=\frac{c}{2}((k_{y}+\imath k_{x})^{3}+(k_{y}-\imath
k_{x})^{3})\boldsymbol\sigma_{z}\ \ .
\end{equation}
This yields:
\begin{equation}
E^{\pm}(k)=\frac{\hbar^{2}k^{2}}{2m^*}\pm\sqrt{(\alpha_R)^{2}k^{2}+c^{2}k^{6}\cos^{2}(3\theta)}\ \ ,
\end{equation}
where $c$ is an anisotropy parameter, and $\theta$ is the in-plane
angle from the $\overline{\Gamma\textmd{M}}$ direction. For small
values of $c$ this expression reduces to the free-electron case
with RB splitting (Eq. (1)). The in-plane asymmetry could be
self-consistently included starting from Eq. (5), but Eq. (7)
provides a minimal alternative model with the single parameter
$c$. The resulting band dispersion agrees well that of the
anisotropic 2DEG proposed by Premper et al. \cite{Premper2007}. It
is shown in Fig. 7 (a) for $|V_{G}|=0$, and (b) for $|V_{G}|=0.3$
eV. The corresponding parameters are summarized in the Appendix.
The model correctly predicts a negative effective mass along
$\overline{\textmd{KMK}}$. Again, the crystal potential induces a
weak RB splitting. The dispersion along
$\overline{\Gamma\textmd{M}\Gamma}$ is essentially identical to
the isotropic case and, as the latter, it overestimates the
spin-splitting of the bands, especially far from
$\overline{\textmd{M}}$ (not shown). The CE contours evolve
continuously with increasing energy from circular to blossom-like,
as in Fig. 7 (c). The sixfold symmetry is the result of the
threefold rotational symmetry and time-reversal symmetry. It
should be noted that the spin polarization has only an in-plane
tangential component for an isotropic 2DEG, whereas a sizeable
out-of-plane and a small radial component are present for the
anisotropic case \cite{Premper2007}. Fig. 8 compares the
experimental CE contours around $\overline{\textmd{M}}$ with the
predictions of the isotropic (left panels) and the anisotropic
(right panels) NFE models. The latter describes reasonably well
the data for $|V_{G}|=0$ but fails to reproduce the dispersion
(Fig. 6 (a)) due to the overestimation of the momentum splitting
at $k$-points far from $\overline{\textmd{M}}$. A finite lattice
potential does open an energy gap, but it perturbs the band
structure and the agreement is completely spoiled.

A closer examination of the CE contours reveals another subtle
inaccuracy of the anisotropic NFE model. The spin polarization
symmetry, determined by the mirror plane of the trimer
configuration \cite{NFE}, is not in agreement with the tip
orientation of the outer CE contour of Fig. 7 (c). This
inconsistency is removed by the tight-binding model considered in
the following subsection.

\subsection{D. An Empirical Tight-Binding Model}

The covalent character of the bonds and the semiconducting nature
of the Bi-Si(111) interface suggest that local orbitals may be a
better starting point. We have performed an empirical tight
binding (TB) calculation which is able to reproduce the main
experimental features. In order to limit the complexity of the
calculation, the model considers a single orbital per atomic site
with $sp_{\textmd{z}}$ symmetry. While this is an approximation,
we expect contributions from other orbital symmetries to be small
in the energy range of interest. This is supported by recent
results for the isostructural Bi-Ge(111) interface
\cite{Hatta2009}. The primitive unit cell contains three Bi atoms,
labelled a, b and c in Fig. 9. In the same figure the five
inequivalent hopping terms are indicated by arrows. All the other
terms can be generated by symmetry. The Si(111) substrate is only
indirectly taken into account through the effective hopping
parameters. A calculation of the transfer integrals is a non
trivial computational task, which clearly goes beyond the scope of
this work. Therefore we defined them in a purely phenomenological
way, assuming an inverse power-law dependence of the distance $d$
between two centers: $V(d)=ad^{-b}$. The prefactor $a$ determines
the bandwidths, while the exponent $b$ determines details of the
dispersion. There is obviously no angular dependence for
$sp_{\textmd{z}}$ states. The Bi-Bi distance within a single
trimer was set to 2.6 \AA, which is very close to literature
values \cite{Wan1992,Gierz2009}. In the actual calculation we
included interactions up to 4$^{\textmd{th}}$ nearest neighbors.
The required overall resemblance with the experimental dispersion
significantly limits the acceptable parameter space. The chosen
values are summarized in Table II of the Appendix.

Figure 10 illustrates the results of the TB model before the
inclusion of a RB interaction. The calculation yields three bands,
corresponding to the three orbitals per unit cell. The two higher
lying states are shown for the $\overline{\Gamma\textmd{M}\Gamma}$
(a) and $\overline{\Gamma\textmd{KM}}$ (b) directions in Fig. 10.
They can be associated to the experimental features of Fig. 4. The
model correctly predicts a double degeneracy at the
$\overline{\Gamma}$ and $\overline{\textmd{K}}$ points,
independent of the parameter values. A double degeneracy is
imposed by the C$_{3\textmd{v}}$ symmetry of these points for the
trimer structure. The $\overline{\textmd{M}}$ point has a lower
symmetry (C$_{1\textmd{h}}$) and therefore no degeneracy is
expected there. An interesting result  is the presence of two
maxima for S$_1$ on both sides of $\overline{\textmd{M}}$ along
$\overline{\Gamma\textmd{M}\Gamma}$. This hallmark of the trimer
structure, which is not observed in simple hexagonal structures,
has been reported for other similar systems
\cite{Kim2009,Lee2008}.  We shall see below that it plays a
potentially important role in the appearance of a $giant$ SO
splitting in Bi-Si(111).

\begin{figure}
  \centering
  \includegraphics[width = 6 cm]{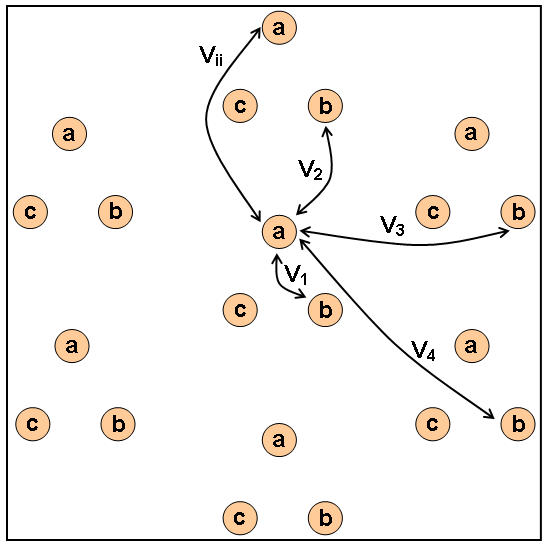}
  \caption{(color online) Schematics of
  the cell used for the TB calculation, with the definition of the inequivalent
  transfer integrals.
  } \label{fig9}
\end{figure}
\begin{figure}
  \centering
  \includegraphics[width = 7.5 cm]{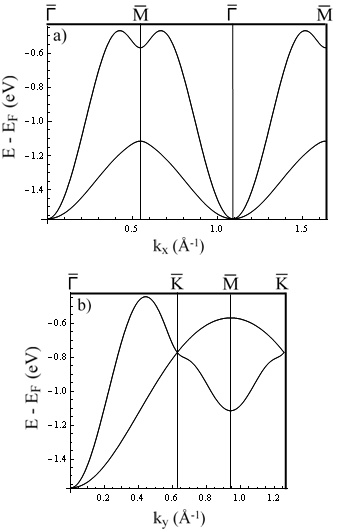}
  \caption{Calculated band dispersion of the two higher lying states according to the TB model without RB interaction,
  along the $\overline{\Gamma\textmd{M}\Gamma}$ (a) and the $\overline{\Gamma\textmd{KM}}$ (b) high-symmetry directions.} \label{fig10}
\end{figure}

We consider next the effect of the SO interaction, by adding a RB term to the TB hamiltonian \cite{Kane2005,Liu2009,Liu22009,Ruegg2009}:
\begin{equation}
H_{\textmd{TB}}=\sum_{<i,j>} V_{ij}c_{is}^{\dagger}c_{js}+\imath
\sum_{<i,j>s,s'}\lambda_{ij}c_{is}^{\dagger}(\mathbf{\boldsymbol\sigma}\times\mathbf{\hat{d}_{ij}})_{z}c_{js'}
\end{equation}
The first term is the usual spin-independent TB hamiltonian, while
the second term is the appropriate TB form for the Rashba
interaction. $c_{is}^{\dagger} (c_{is})$ is the creation
(annihilation) operator of an electron with spin $s$ ($\uparrow$
or $\downarrow$) on atomic site $i$, $V_{ij}$ and $\lambda_{ij}$
are the transfer integrals and the SO coefficients. The latter are
generated by a similar power-law function of the distance, but
with independent parameters (see Table II of the Appendix).
$\boldsymbol\sigma$ is the vector of the Pauli matrices and
$\mathbf{\hat{d}_{ij}}$ is the vector connecting site $j$ to site
$i$.

The six basis vectors of a site-spin representation
are, with obvious notation,
$|a\uparrow\rangle$, $|b\uparrow\rangle$, $|c\uparrow\rangle$,
$|a\downarrow\rangle$, $|b\downarrow\rangle$,
$|c\downarrow\rangle$. In this representation, the hamiltonian
matrix has the form:
\begin{equation}
H_{\textmd{TB}}=
\left(\begin{array}{cc}
           H_{0}& H_{\textmd{R}} \\
           H_{\textmd{R}}^{*}&H_{0} \\
\end{array}\right)\ \ .
\end{equation}
$H_{0}$ is the $3\times3$ spin-independent TB hamiltonian which
describes states of equal spin. $H_{\textmd{R}}$ is a $3\times3$
matrix generated by the second term of $H_{\textmd{TB}}$, which
describes the coupling of electrons with opposite spins. Our
method is essentially akin to the TB model of Ref.
\onlinecite{Petersen2000} and can qualitatively describe the SO
split bands, namely their topology.

\begin{figure}
  \centering
  \includegraphics[width = 7.5 cm]{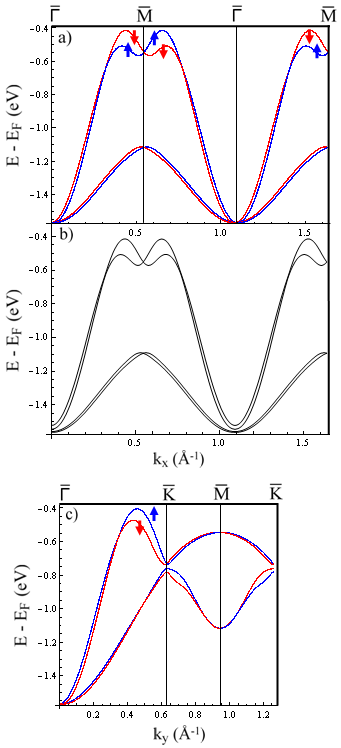}
  \caption{(color online) Calculated bands for the TB model with SO (see Table II of the Appendix for the parameters) for:
(a) the $\overline{\Gamma\textmd{M}\Gamma}$ direction; (b) the
parallel cut $b$ of Fig. 4 (d) ($k_\textmd{{y}}=0.05$ \AA$^{-1}$)
(b); (c) the $\overline{\Gamma\textmd{KM}}$ direction. Arrows
denote the main component of the spin polarization of the SO-split
branches. The polarization difference is not 100\% due to
additional radial and out-of-plane components.} \label{fig11}
\end{figure}

Figure 11 shows the band structure of the TB model with RB
interaction, for the parameters which best reproduce the
experimental results \cite{TBAu}. The first obvious effect of the
RB interaction is that all states are now split. Both S$_{1}$ and
S$_{2}$ are doubly-degenerate at $\overline{\Gamma}$, and exhibit
an isotropic dispersion around this point (Fig. 11 (a)). By moving
out of the high-symmetry line (Fig. 11 (b)) the degeneracy is
lifted, as expected for the usual RB scenario of Fig. 1. The
spin-split branches cross again at $\overline{\textmd{M}}$, as
required by time-reversal symmetry. Here, the dispersion of S$_1$
is strongly anisotropic. The splitting of S$_1$  along the
$\overline{\Gamma\textmd{M}\Gamma}$ direction is much larger than
that of S$_2$. It is also much larger than the splitting of S$_1$
around $\overline{\Gamma}$. By contrast, it is small along the SBZ
boundary $\overline{\textmd{KMK}}$ (Fig. 11 (c)). All these
features of the band structure agree with the ARPES results, and
also with the results of first-principles calculations
\cite{Gierz2009,Sakamoto2009}. The calculated energy difference
E$_R(\overline{\textmd{M}})$ between the band crossing at
$\overline{\textmd{M}}$ and the band maximum is nonetheless
smaller than the experimental one.

A comparison of Fig. 10 (a) and Fig. 11 (a) shows that the large
SO splitting of S$_1$ is a consequence of the split maxima on
opposite sides of $\overline{\textmd{M}}$, rather than the result
of a large SO coupling. In a way, the conditions for a large
momentum splitting are already present in the band structure, and
the main effect of the SO interaction is to split in energy the
two subbands. Indeed for S$_2$, which has a maximum at
$\overline{\textmd{M}}$ in the parent structure, the Rashba
splitting is small. The origin of the large momentum separation is
therefore rather different from that of the `giant' splitting
observed at metallic interfaces like BiAg$_{2}$/Ag(111)
\cite{Ast2007}. In the TB framework, it is due to the second term
of the Hamiltonian (Eq. (10)), i.e. the way the trimer arrangement
determines the hybridization of Bi orbitals with unlike spins.

\begin{figure}
  \centering
  \includegraphics[width = 8.5 cm]{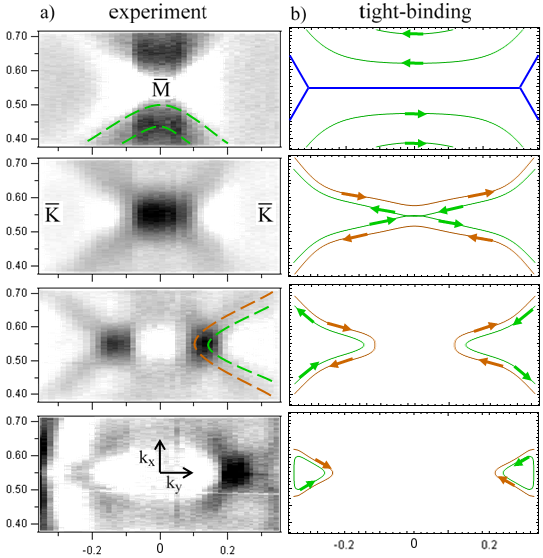}
  \caption{(color online) (a) CE ARPES contours. The dashed curves on the experimental maps are guides to the eye. Green (light gray) and brown (dark gray) colors indicate positive and negative values of tangential spin polarization. (b) CE contours for the TB model. The arrows indicate the
  in-plane projection of the spin polarization. The energies were adjusted to correspond to the experimental values.} \label{fig12}
\end{figure}

Figure 12 illustrates the energy evolution of the calculated CE
contours near $\overline{\textmd{M}}$, across the spin-degeneracy
point. The experimental contours are reproduced here for a
qualitative comparison. A fully quantitative comparison is not
possible due to the already mentioned difference in
E$_R(\overline{\textmd{M}})$. The energies of the calculated
contours were therefore adjusted to correspond to the experimental
energies of the ARPES contours. The model yields open contours
around $\overline{\textmd{M}}$, which are in reasonable agreement
with the topology of the experimental bands. The CE contours are
shown on a broader momentum range in Fig. 13 (a) and (b) for two
energies above, and in Fig. 13 (c) for an energy well below the
crossing point. The shape of the contours is nearly circular near
the bottom of the band at $\overline{\Gamma}$ (Fig. 13 (c)), and
it evolves to a hexagonal and finally blossom-like shape at larger
energy. This is seen more clearly in Fig. 13 (d), where the SO
parameters were artificially increased to enhance the splitting of
the two subbands. Interestingly, the blossom-like shape of Fig. 13
(d) is identical to the one predicted by the anisotropic NFE model
(see subsection III (C) and Ref. \cite{Premper2007} but the tips
now correctly point along the $\overline{\Gamma\textmd{M}\Gamma}$
direction.

\begin{figure}
  \centering
  \includegraphics[width = 7.2 cm]{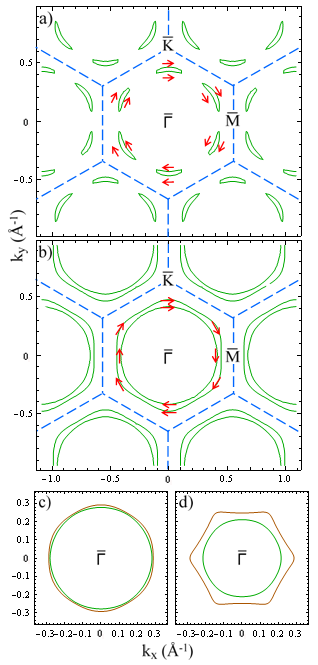}
  \caption{(color online) CE contours of the TB model for energies above (a), (b) and well below (c) the spin degenerate point at $\overline{\textmd{M}}$. Arrows indicate the in-plane projection of the spin
  polarization. (d) is the same as (c) for an eightfold increase in the SO
  parameters. Green (light gray) and brown (dark gray) colors indicate positive and negative values of tangential spin polarization.
  } \label{fig13}
\end{figure}

\section{IV. Conclusions and Outlook}

We performed a detailed ARPES study of the SO-split electronic
states in the $1$ ML trimer phase of the Bi-Si(111) interface. We
paid special attention to the region of $k$-space close to the
$\overline{\textmd{M}}$ point, where the topmost S$_1$ hybrid
surface state is both not degenerate with bulk states and distinct
from other surface-related features. This region is of particular
interest because there S$_1$ exhibits a large and non-conventional
Rashba-like splitting. Energy-dependent constant energy contours
clarify the complex topology of the SO-split states, and underline
the differences with a standard RB scenario. The ARPES data show
that the interface has an insulating character, but the Fermi
level could be moved into the SO-split bands by applying an
external electric potential in a back-gated structure. It should
then be possible to advantageously exploit the large momentum
separation of the two spin-polarized subbands in a spin
field-effect transistor \cite{Datta1990,Koo2009}.

We have used the predictions of three simple models for a 2DEG in
the presence of SO interaction as guidelines for the
interpretation of the experimental results. The comparison of the
NFE and local-orbital schemes, which proceed from opposite
starting points, has a certain didactic value. Moreover we were
able to assess the limits of the various approaches applied to the
Bi-Si(111) case. The ARPES results and CE contours are well
described by  a NFE RB effect in a sufficiently small region
around the $\overline{\Gamma}$ point. Further away from
$\overline{\Gamma}$ the isotropic NFE model must be refined to
include an in-plane asymmetry. Near the SZB, the specific symmetry
properties of the interface determine the characteristics of the
SO splitting \cite{Oguchi2009}, and only the empirical TB model
captures the salient features of the electronic structure. The
same model predicts a peculiar spin texture (Fig. 13 (a)). Hole
pockets with a non-vortical spin arrangement are reminiscent of
the teardrop Fermi surface contours of the topological insulator
Bi$_{\textmd{1-x}}$Sb$_{\textmd{x}}$ \cite{Hsieh2009}. Within a
few meV the pockets develop into two connected concentric contours
with the same spin polarization (Fig. 13 (b)). This new prediction
calls for an experimental verification by spin-resolved ARPES.

Our TB approximation obviously cannot reproduce all the details of
the complicated electronic structure. More elaborate TB schemes
could be implemented by extending the set of local orbitals, by
treating in an explicit way the hybridization with the substrate,
and by deriving the relevant transfer integrals from a direct
calculation. However, the actual merit of such schemes would be
dubious, since the computational complexity would approach that of
first principles calculations, and since the immediate simplicity
of the model would be lost.

The main insight from our analysis of the new experimental data is
the realization that the $giant$ SO splitting at this interface is
not primarily controlled by a large atomic SO parameter, as in the
case of BiAg$_{2}$/Ag(111) and other metallic surface alloys
\cite{Ast2007}. On the contrary, the effect is largely due to a
peculiar feature in the band structure, namely the presence of
symmetrically split maxima around the $\overline{\textmd{M}}$
point. This is a new, unexpected mechanism to achieve large spin
separation at an interface. The underlying band feature is
characteristic of the trimer structure, and it has been identified
at other similar interfaces \cite{Kim2009,Lee2008}. We may
therefore anticipate that similar large "Rashba"-like effects
could be discovered in other systems characterized by moderate SO
parameters. \\

\begin{acknowledgments}
It is a pleasure to acknowledge fruitful discussions with A.
Baldereschi, J. Henk, I. Rousochatzakis and K. Sakamoto. We also
thank L. Moreschini and L. Casanellas for help during the early
stage of this work. E.F. acknowledges the Alexander S. Onassis
Public Benefit Foundation for the award of a scholarship. This
research was supported by the Swiss NSF and the NCCR MaNEP.
\end{acknowledgments}

\section{Appendix}

The phenomenological parameters of the models are summarized in the
following tables:

\begin{table}[h!tb]
\caption{Parameters of the NFE modes. The potential value refers to the 1st Fourier
coefficient of the crystal potential V(r).}
\begin{tabular}{c c c}
  \hline \hline
  parameter & isotropic NFE & anisotropic NFE \\
  \hline
  m$^{*}$ (m$_{e}$) & 0.8 & 0.8 \\
  c (eV.\AA$^{3}$) & 0 & 4.8 \\
  $\alpha_{R}$ (eV.\AA) & 1.1 & 1.1 \\
  $|V_{G}|$ (eV) & 0.3 & 0.3 \\
  \hline \hline
\end{tabular}
\end{table}

\begin{table}[h!tb]
\caption{Parameters of the TB model. The  hopping and SO parameters are
generated by power law functions of the distance $d$ (i.e. $a
d^{-b}$).}
\begin{tabular}{c c c c c}
  \hline \hline
  &  &\ \   $V_{ij}$ &\ \  & $\lambda_{ij}$\\
  \hline
  prefactor $a$ &\ \   & -2.94 &\ \   & 0.15 (1.20 in Fig. 13 (d)) \\
  exponent $b$ &\ \   & 1.13 &\ \   & 0.80 \\
  \hline \hline
\end{tabular}
\end{table}


\begin{thebibliography}{48}
\expandafter\ifx\csname
natexlab\endcsname\relax\def\natexlab#1{#1}\fi
\expandafter\ifx\csname bibnamefont\endcsname\relax
  \def\bibnamefont#1{#1}\fi
\expandafter\ifx\csname bibfnamefont\endcsname\relax
  \def\bibfnamefont#1{#1}\fi
\expandafter\ifx\csname citenamefont\endcsname\relax
  \def\citenamefont#1{#1}\fi
\expandafter\ifx\csname url\endcsname\relax
  \def\url#1{\texttt{#1}}\fi
\expandafter\ifx\csname
urlprefix\endcsname\relax\def\urlprefix{URL }\fi
\providecommand{\bibinfo}[2]{#2}
\providecommand{\eprint}[2][]{\url{#2}}

\bibitem[{\citenamefont{Dresselhaus}(1955)}]{Dresselhaus1955}
\bibinfo{author}{\bibfnamefont{G.}~\bibnamefont{Dresselhaus}},
  \bibinfo{journal}{Phys. Rev.} \textbf{\bibinfo{volume}{100}},
  \bibinfo{pages}{580} (\bibinfo{year}{1955}).

\bibitem[{\citenamefont{K\"{o}nemann et~al.}(2005)\citenamefont{K\"{o}nemann,
  Haug, Maude, Fal'ko, and Altshuler}}]{Konemann2005}
\bibinfo{author}{\bibfnamefont{J.}~\bibnamefont{K\"{o}nemann}},
  \bibinfo{author}{\bibfnamefont{R.~J.} \bibnamefont{Haug}},
  \bibinfo{author}{\bibfnamefont{D.~K.} \bibnamefont{Maude}},
  \bibinfo{author}{\bibfnamefont{V.~I.} \bibnamefont{Fal'ko}},
  \bibnamefont{and} \bibinfo{author}{\bibfnamefont{B.~L.}
  \bibnamefont{Altshuler}}, \bibinfo{journal}{Phys. Rev. Lett.}
  \textbf{\bibinfo{volume}{94}}, \bibinfo{pages}{226404}
  (\bibinfo{year}{2005}).

\bibitem[{\citenamefont{Bychkov and Rashba}(1984)}]{Bychkov1984}
\bibinfo{author}{\bibfnamefont{Y.~A.} \bibnamefont{Bychkov}} \bibnamefont{and}
  \bibinfo{author}{\bibfnamefont{E.~I.} \bibnamefont{Rashba}},
  \bibinfo{journal}{JETP Lett.} \textbf{\bibinfo{volume}{39}},
  \bibinfo{pages}{78} (\bibinfo{year}{1984}).

\bibitem[{\citenamefont{LaShell et~al.}(1996)\citenamefont{LaShell, McDougall,
  and Jensen}}]{LaShell1996}
\bibinfo{author}{\bibfnamefont{S.}~\bibnamefont{LaShell}},
  \bibinfo{author}{\bibfnamefont{B.}~\bibnamefont{McDougall}},
  \bibnamefont{and} \bibinfo{author}{\bibfnamefont{E.}~\bibnamefont{Jensen}},
  \bibinfo{journal}{Phys. Rev. Lett.} \textbf{\bibinfo{volume}{77}},
  \bibinfo{pages}{3419} (\bibinfo{year}{1996}).

\bibitem[{\citenamefont{Rotenberg et~al.}(1999)\citenamefont{Rotenberg, Chung,
  and Kevan}}]{Rotenberg1999}
\bibinfo{author}{\bibfnamefont{E.}~\bibnamefont{Rotenberg}},
  \bibinfo{author}{\bibfnamefont{J.~W.} \bibnamefont{Chung}}, \bibnamefont{and}
  \bibinfo{author}{\bibfnamefont{S.~D.} \bibnamefont{Kevan}},
  \bibinfo{journal}{Phys. Rev. Lett.} \textbf{\bibinfo{volume}{82}},
  \bibinfo{pages}{4066} (\bibinfo{year}{1999}).

\bibitem[{\citenamefont{Hochstrasser et~al.}(2002)\citenamefont{Hochstrasser,
  Tobin, Rotenberg, and Kevan}}]{Hochstrasser2002}
\bibinfo{author}{\bibfnamefont{M.}~\bibnamefont{Hochstrasser}},
  \bibinfo{author}{\bibfnamefont{J.~G.} \bibnamefont{Tobin}},
  \bibinfo{author}{\bibfnamefont{E.}~\bibnamefont{Rotenberg}},
  \bibnamefont{and} \bibinfo{author}{\bibfnamefont{S.~D.} \bibnamefont{Kevan}},
  \bibinfo{journal}{Phys. Rev. Lett.} \textbf{\bibinfo{volume}{89}},
  \bibinfo{pages}{216802} (\bibinfo{year}{2002}).

\bibitem[{\citenamefont{Reinert}(2003)}]{Reinert2003}
\bibinfo{author}{\bibfnamefont{F.}~\bibnamefont{Reinert}}, \bibinfo{journal}{J.
  Phys. Condens. Matter} \textbf{\bibinfo{volume}{15}}, \bibinfo{pages}{S693}
  (\bibinfo{year}{2003}).

\bibitem[{\citenamefont{Koroteev et~al.}(2004)\citenamefont{Koroteev,
  Bihlmayer, Gayone, Chulkov, Bl\"{u}gel, Echenique, and
  Hofmann}}]{Koroteev2004}
\bibinfo{author}{\bibfnamefont{Y.~M.} \bibnamefont{Koroteev}},
  \bibinfo{author}{\bibfnamefont{G.}~\bibnamefont{Bihlmayer}},
  \bibinfo{author}{\bibfnamefont{J.~E.} \bibnamefont{Gayone}},
  \bibinfo{author}{\bibfnamefont{E.~V.} \bibnamefont{Chulkov}},
  \bibinfo{author}{\bibfnamefont{S.}~\bibnamefont{Bl\"{u}gel}},
  \bibinfo{author}{\bibfnamefont{P.~M.} \bibnamefont{Echenique}},
  \bibnamefont{and} \bibinfo{author}{\bibfnamefont{P.}~\bibnamefont{Hofmann}},
  \bibinfo{journal}{Phys. Rev. Lett.} \textbf{\bibinfo{volume}{93}},
  \bibinfo{pages}{046403} (\bibinfo{year}{2004}).

\bibitem[{\citenamefont{Henk et~al.}(2004)\citenamefont{Henk, Hoesch,
  Osterwalder, Ernst, and Bruno}}]{Henk2004}
\bibinfo{author}{\bibfnamefont{J.}~\bibnamefont{Henk}},
  \bibinfo{author}{\bibfnamefont{M.}~\bibnamefont{Hoesch}},
  \bibinfo{author}{\bibfnamefont{J.}~\bibnamefont{Osterwalder}},
  \bibinfo{author}{\bibfnamefont{A.}~\bibnamefont{Ernst}}, \bibnamefont{and}
  \bibinfo{author}{\bibfnamefont{P.}~\bibnamefont{Bruno}}, \bibinfo{journal}{J.
  Phys. Condens. Matter} \textbf{\bibinfo{volume}{16}}, \bibinfo{pages}{7581}
  (\bibinfo{year}{2004}).

\bibitem[{\citenamefont{Malterre et~al.}(2007)\citenamefont{Malterre, Kierren,
  Fagot-Revurat, Pons, Tejeda, Didiot, Cercellier, and
  Bendounan}}]{Malterre2007}
\bibinfo{author}{\bibfnamefont{D.}~\bibnamefont{Malterre}},
  \bibinfo{author}{\bibfnamefont{B.}~\bibnamefont{Kierren}},
  \bibinfo{author}{\bibfnamefont{Y.}~\bibnamefont{Fagot-Revurat}},
  \bibinfo{author}{\bibfnamefont{S.}~\bibnamefont{Pons}},
  \bibinfo{author}{\bibfnamefont{A.}~\bibnamefont{Tejeda}},
  \bibinfo{author}{\bibfnamefont{C.}~\bibnamefont{Didiot}},
  \bibinfo{author}{\bibfnamefont{H.}~\bibnamefont{Cercellier}},
  \bibnamefont{and}
  \bibinfo{author}{\bibfnamefont{A.}~\bibnamefont{Bendounan}},
  \bibinfo{journal}{New. J. Phys.} \textbf{\bibinfo{volume}{9}},
  \bibinfo{pages}{391} (\bibinfo{year}{2007}).

\bibitem[{\citenamefont{Petersen and Hedegard}(2000)}]{Petersen2000}
\bibinfo{author}{\bibfnamefont{L.}~\bibnamefont{Petersen}} \bibnamefont{and}
  \bibinfo{author}{\bibfnamefont{P.}~\bibnamefont{Hedegard}},
  \bibinfo{journal}{Surf. Sci.} \textbf{\bibinfo{volume}{459}},
  \bibinfo{pages}{49} (\bibinfo{year}{2000}).

\bibitem[{\citenamefont{Bihlmayer et~al.}(2006)\citenamefont{Bihlmayer,
  Koroteev, Echenique, Chulkov, and Bl\"{u}gel}}]{Bihlmayer2006}
\bibinfo{author}{\bibfnamefont{G.}~\bibnamefont{Bihlmayer}},
  \bibinfo{author}{\bibfnamefont{Y.~M.} \bibnamefont{Koroteev}},
  \bibinfo{author}{\bibfnamefont{P.~M.} \bibnamefont{Echenique}},
  \bibinfo{author}{\bibfnamefont{E.~V.} \bibnamefont{Chulkov}},
  \bibnamefont{and}
  \bibinfo{author}{\bibfnamefont{S.}~\bibnamefont{Bl\"{u}gel}},
  \bibinfo{journal}{Surf. Sci.} \textbf{\bibinfo{volume}{600}},
  \bibinfo{pages}{3888} (\bibinfo{year}{2006}).

\bibitem[{\citenamefont{Nagano et~al.}(2009)\citenamefont{Nagano, Kodama,
  Shishidou, and Oguchi}}]{Nagano2009}
\bibinfo{author}{\bibfnamefont{M.}~\bibnamefont{Nagano}},
  \bibinfo{author}{\bibfnamefont{A.}~\bibnamefont{Kodama}},
  \bibinfo{author}{\bibfnamefont{T.}~\bibnamefont{Shishidou}},
  \bibnamefont{and} \bibinfo{author}{\bibfnamefont{T.}~\bibnamefont{Oguchi}},
  \bibinfo{journal}{J. Phys.: Condens. Matter} \textbf{\bibinfo{volume}{21}},
  \bibinfo{pages}{064239} (\bibinfo{year}{2009}).

\bibitem[{\citenamefont{Ast et~al.}(2007)\citenamefont{Ast, Henk, Ernst,
  Moreschini, Falub, Pacil\'{e}, Bruno, Kern, and Grioni}}]{Ast2007}
\bibinfo{author}{\bibfnamefont{C.~R.} \bibnamefont{Ast}},
  \bibinfo{author}{\bibfnamefont{J.}~\bibnamefont{Henk}},
  \bibinfo{author}{\bibfnamefont{A.}~\bibnamefont{Ernst}},
  \bibinfo{author}{\bibfnamefont{L.}~\bibnamefont{Moreschini}},
  \bibinfo{author}{\bibfnamefont{M.~C.} \bibnamefont{Falub}},
  \bibinfo{author}{\bibfnamefont{D.}~\bibnamefont{Pacil\'{e}}},
  \bibinfo{author}{\bibfnamefont{P.}~\bibnamefont{Bruno}},
  \bibinfo{author}{\bibfnamefont{K.}~\bibnamefont{Kern}}, \bibnamefont{and}
  \bibinfo{author}{\bibfnamefont{M.}~\bibnamefont{Grioni}},
  \bibinfo{journal}{Phys. Rev. Lett.} \textbf{\bibinfo{volume}{98}},
  \bibinfo{pages}{186807} (\bibinfo{year}{2007}).

\bibitem[{\citenamefont{Meier et~al.}(2008)\citenamefont{Meier, Dil,
  Lobo-Checa, Patthey, and Osterwalder}}]{Meier2008}
\bibinfo{author}{\bibfnamefont{F.}~\bibnamefont{Meier}},
  \bibinfo{author}{\bibfnamefont{H.}~\bibnamefont{Dil}},
  \bibinfo{author}{\bibfnamefont{J.}~\bibnamefont{Lobo-Checa}},
  \bibinfo{author}{\bibfnamefont{L.}~\bibnamefont{Patthey}}, \bibnamefont{and}
  \bibinfo{author}{\bibfnamefont{J.}~\bibnamefont{Osterwalder}},
  \bibinfo{journal}{Phys. Rev. B} \textbf{\bibinfo{volume}{77}},
  \bibinfo{pages}{165431} (\bibinfo{year}{2008}).

\bibitem[{\citenamefont{Premper et~al.}(2007)\citenamefont{Premper, Trautmann,
  Henk, and Bruno}}]{Premper2007}
\bibinfo{author}{\bibfnamefont{J.}~\bibnamefont{Premper}},
  \bibinfo{author}{\bibfnamefont{M.}~\bibnamefont{Trautmann}},
  \bibinfo{author}{\bibfnamefont{J.}~\bibnamefont{Henk}}, \bibnamefont{and}
  \bibinfo{author}{\bibfnamefont{P.}~\bibnamefont{Bruno}},
  \bibinfo{journal}{Phys. Rev. B.} \textbf{\bibinfo{volume}{76}},
  \bibinfo{pages}{073310} (\bibinfo{year}{2007}).

\bibitem[{\citenamefont{Bihlmayer et~al.}(2007)\citenamefont{Bihlmayer,
  Bl\"{u}gel, and Chulkov}}]{Bihlmayer2007}
\bibinfo{author}{\bibfnamefont{G.}~\bibnamefont{Bihlmayer}},
  \bibinfo{author}{\bibfnamefont{S.}~\bibnamefont{Bl\"{u}gel}},
  \bibnamefont{and} \bibinfo{author}{\bibfnamefont{E.~V.}
  \bibnamefont{Chulkov}}, \bibinfo{journal}{Phys. Rev. B}
  \textbf{\bibinfo{volume}{75}}, \bibinfo{pages}{195414}
  (\bibinfo{year}{2007}).

\bibitem[{\citenamefont{Datta and Das}(1990)}]{Datta1990}
\bibinfo{author}{\bibfnamefont{S.}~\bibnamefont{Datta}} \bibnamefont{and}
  \bibinfo{author}{\bibfnamefont{B.}~\bibnamefont{Das}},
  \bibinfo{journal}{Appl. Phys. Lett.} \textbf{\bibinfo{volume}{56}},
  \bibinfo{pages}{665} (\bibinfo{year}{1990}).

\bibitem[{\citenamefont{Koo et~al.}(2009)\citenamefont{Koo, Kwon, Eom, Chang,
  Han, and Johnson}}]{Koo2009}
\bibinfo{author}{\bibfnamefont{H.~C.} \bibnamefont{Koo}},
  \bibinfo{author}{\bibfnamefont{J.~H.} \bibnamefont{Kwon}},
  \bibinfo{author}{\bibfnamefont{J.}~\bibnamefont{Eom}},
  \bibinfo{author}{\bibfnamefont{J.}~\bibnamefont{Chang}},
  \bibinfo{author}{\bibfnamefont{S.~H.} \bibnamefont{Han}}, \bibnamefont{and}
  \bibinfo{author}{\bibfnamefont{M.}~\bibnamefont{Johnson}},
  \bibinfo{journal}{Science} \textbf{\bibinfo{volume}{325}},
  \bibinfo{pages}{1515} (\bibinfo{year}{2009}).

\bibitem[{\citenamefont{Frantzeskakis et~al.}(2008)\citenamefont{Frantzeskakis,
  Pons, Mirhosseini, Henk, Ast, and Grioni}}]{Frantzeskakis2008}
\bibinfo{author}{\bibfnamefont{E.}~\bibnamefont{Frantzeskakis}},
  \bibinfo{author}{\bibfnamefont{S.}~\bibnamefont{Pons}},
  \bibinfo{author}{\bibfnamefont{H.}~\bibnamefont{Mirhosseini}},
  \bibinfo{author}{\bibfnamefont{J.}~\bibnamefont{Henk}},
  \bibinfo{author}{\bibfnamefont{C.~R.} \bibnamefont{Ast}}, \bibnamefont{and}
  \bibinfo{author}{\bibfnamefont{M.}~\bibnamefont{Grioni}},
  \bibinfo{journal}{Phys. Rev. Lett.} \textbf{\bibinfo{volume}{101}},
  \bibinfo{pages}{196805} (\bibinfo{year}{2008}).

\bibitem[{\citenamefont{Hirahara et~al.}(2006)\citenamefont{Hirahara, Nagao,
  Matsuda, Bihlmayer, Chulkov, Koroteev, Echenique, Saito, and
  Hasegawa}}]{Hirahara2006}
\bibinfo{author}{\bibfnamefont{T.}~\bibnamefont{Hirahara}},
  \bibinfo{author}{\bibfnamefont{T.}~\bibnamefont{Nagao}},
  \bibinfo{author}{\bibfnamefont{I.}~\bibnamefont{Matsuda}},
  \bibinfo{author}{\bibfnamefont{G.}~\bibnamefont{Bihlmayer}},
  \bibinfo{author}{\bibfnamefont{E.~V.} \bibnamefont{Chulkov}},
  \bibinfo{author}{\bibfnamefont{Y.~M.} \bibnamefont{Koroteev}},
  \bibinfo{author}{\bibfnamefont{P.~M.} \bibnamefont{Echenique}},
  \bibinfo{author}{\bibfnamefont{M.}~\bibnamefont{Saito}}, \bibnamefont{and}
  \bibinfo{author}{\bibfnamefont{S.}~\bibnamefont{Hasegawa}},
  \bibinfo{journal}{Phys. Rev. Lett.} \textbf{\bibinfo{volume}{97}},
  \bibinfo{pages}{146803} (\bibinfo{year}{2006}).

\bibitem[{\citenamefont{He et~al.}(2008)\citenamefont{He, Hirahara, Okuda,
  Hasegawa, Kazizaki, and Matsuda}}]{He2008}
\bibinfo{author}{\bibfnamefont{K.}~\bibnamefont{He}},
  \bibinfo{author}{\bibfnamefont{T.}~\bibnamefont{Hirahara}},
  \bibinfo{author}{\bibfnamefont{T.}~\bibnamefont{Okuda}},
  \bibinfo{author}{\bibfnamefont{S.}~\bibnamefont{Hasegawa}},
  \bibinfo{author}{\bibfnamefont{A.}~\bibnamefont{Kazizaki}}, \bibnamefont{and}
  \bibinfo{author}{\bibfnamefont{I.}~\bibnamefont{Matsuda}},
  \bibinfo{journal}{Phys. Rev. Lett.} \textbf{\bibinfo{volume}{101}},
  \bibinfo{pages}{107604} (\bibinfo{year}{2008}).

\bibitem[{\citenamefont{Frantzeskakis et~al.}(2010)\citenamefont{Frantzeskakis,
  Crepaldi, Pons, Kern, and Grioni}}]{FrantzeskakisJElSpectr2010}
\bibinfo{author}{\bibfnamefont{E.}~\bibnamefont{Frantzeskakis}},
  \bibinfo{author}{\bibfnamefont{A.}~\bibnamefont{Crepaldi}},
  \bibinfo{author}{\bibfnamefont{S.}~\bibnamefont{Pons}},
  \bibinfo{author}{\bibfnamefont{K.}~\bibnamefont{Kern}}, \bibnamefont{and}
  \bibinfo{author}{\bibfnamefont{M.}~\bibnamefont{Grioni}},
  \bibinfo{journal}{J. Electron Spectr. Rel. Phenom.}
  \textbf{\bibinfo{volume}{181}}, \bibinfo{pages}{88} (\bibinfo{year}{2010}).

\bibitem[{\citenamefont{Dil et~al.}(2008)\citenamefont{Dil, Meier, Lobo-Checa,
  Patthey, Bihlmayer, and Osterwalder}}]{Dil2008}
\bibinfo{author}{\bibfnamefont{J.~H.} \bibnamefont{Dil}},
  \bibinfo{author}{\bibfnamefont{F.}~\bibnamefont{Meier}},
  \bibinfo{author}{\bibfnamefont{J.}~\bibnamefont{Lobo-Checa}},
  \bibinfo{author}{\bibfnamefont{L.}~\bibnamefont{Patthey}},
  \bibinfo{author}{\bibfnamefont{G.}~\bibnamefont{Bihlmayer}},
  \bibnamefont{and}
  \bibinfo{author}{\bibfnamefont{J.}~\bibnamefont{Osterwalder}},
  \bibinfo{journal}{Phys. Rev Lett.} \textbf{\bibinfo{volume}{101}},
  \bibinfo{pages}{266802} (\bibinfo{year}{2008}).

\bibitem[{\citenamefont{Gierz et~al.}(2009)\citenamefont{Gierz, Suzuki,
  Frantzeskakis, Pons, Ostanin, Ernst, Henk, Grioni, Kern, and
  Ast}}]{Gierz2009}
\bibinfo{author}{\bibfnamefont{I.}~\bibnamefont{Gierz}},
  \bibinfo{author}{\bibfnamefont{T.}~\bibnamefont{Suzuki}},
  \bibinfo{author}{\bibfnamefont{E.}~\bibnamefont{Frantzeskakis}},
  \bibinfo{author}{\bibfnamefont{S.}~\bibnamefont{Pons}},
  \bibinfo{author}{\bibfnamefont{S.}~\bibnamefont{Ostanin}},
  \bibinfo{author}{\bibfnamefont{A.}~\bibnamefont{Ernst}},
  \bibinfo{author}{\bibfnamefont{J.}~\bibnamefont{Henk}},
  \bibinfo{author}{\bibfnamefont{M.}~\bibnamefont{Grioni}},
  \bibinfo{author}{\bibfnamefont{K.}~\bibnamefont{Kern}}, \bibnamefont{and}
  \bibinfo{author}{\bibfnamefont{C.~R.} \bibnamefont{Ast}},
  \bibinfo{journal}{Phys. Rev. Lett.} \textbf{\bibinfo{volume}{103}},
  \bibinfo{pages}{046803} (\bibinfo{year}{2009}).

\bibitem[{\citenamefont{Sakamoto et~al.}(2009)\citenamefont{Sakamoto, Kakuta,
  Sugawara, Miyamoto, Kimura, Kuzumaki, Ueno, Annese, Fujii, Kodama
  et~al.}}]{Sakamoto2009}
\bibinfo{author}{\bibfnamefont{K.}~\bibnamefont{Sakamoto}},
  \bibinfo{author}{\bibfnamefont{H.}~\bibnamefont{Kakuta}},
  \bibinfo{author}{\bibfnamefont{K.}~\bibnamefont{Sugawara}},
  \bibinfo{author}{\bibfnamefont{K.}~\bibnamefont{Miyamoto}},
  \bibinfo{author}{\bibfnamefont{A.}~\bibnamefont{Kimura}},
  \bibinfo{author}{\bibfnamefont{T.}~\bibnamefont{Kuzumaki}},
  \bibinfo{author}{\bibfnamefont{N.}~\bibnamefont{Ueno}},
  \bibinfo{author}{\bibfnamefont{E.}~\bibnamefont{Annese}},
  \bibinfo{author}{\bibfnamefont{J.}~\bibnamefont{Fujii}},
  \bibinfo{author}{\bibfnamefont{A.}~\bibnamefont{Kodama}},
  \bibnamefont{et~al.}, \bibinfo{journal}{Phys. Rev. Lett.}
  \textbf{\bibinfo{volume}{103}}, \bibinfo{pages}{156801}
  (\bibinfo{year}{2009}).

\bibitem[{\citenamefont{Hatta et~al.}(2009)\citenamefont{Hatta, Aruga, Ohtsubo,
  and Okuyama}}]{Hatta2009}
\bibinfo{author}{\bibfnamefont{S.}~\bibnamefont{Hatta}},
  \bibinfo{author}{\bibfnamefont{T.}~\bibnamefont{Aruga}},
  \bibinfo{author}{\bibfnamefont{Y.}~\bibnamefont{Ohtsubo}}, \bibnamefont{and}
  \bibinfo{author}{\bibfnamefont{H.}~\bibnamefont{Okuyama}},
  \bibinfo{journal}{Phys. Rev. B} \textbf{\bibinfo{volume}{80}},
  \bibinfo{pages}{113309} (\bibinfo{year}{2009}).

\bibitem[{\citenamefont{Yaji et~al.}(2010)\citenamefont{Yaji, Ohtsubo, Hatta,
  Okuyama, Miyamoto, Okuda, Kimura, Namatame, Taniguchi, and Aruga}}]{Yaji2010}
\bibinfo{author}{\bibfnamefont{K.}~\bibnamefont{Yaji}},
  \bibinfo{author}{\bibfnamefont{Y.}~\bibnamefont{Ohtsubo}},
  \bibinfo{author}{\bibfnamefont{S.}~\bibnamefont{Hatta}},
  \bibinfo{author}{\bibfnamefont{H.}~\bibnamefont{Okuyama}},
  \bibinfo{author}{\bibfnamefont{K.}~\bibnamefont{Miyamoto}},
  \bibinfo{author}{\bibfnamefont{T.}~\bibnamefont{Okuda}},
  \bibinfo{author}{\bibfnamefont{A.}~\bibnamefont{Kimura}},
  \bibinfo{author}{\bibfnamefont{H.}~\bibnamefont{Namatame}},
  \bibinfo{author}{\bibfnamefont{M.}~\bibnamefont{Taniguchi}},
  \bibnamefont{and} \bibinfo{author}{\bibfnamefont{T.}~\bibnamefont{Aruga}},
  \bibinfo{journal}{Nature Communications} \textbf{\bibinfo{volume}{1}},
  \bibinfo{pages}{17} (\bibinfo{year}{2010}).

\bibitem[{\citenamefont{Nakatani et~al.}(1995)\citenamefont{Nakatani,
  Takahashi, Kuwahara, and Aono}}]{Nakatani1995}
\bibinfo{author}{\bibfnamefont{S.}~\bibnamefont{Nakatani}},
  \bibinfo{author}{\bibfnamefont{T.}~\bibnamefont{Takahashi}},
  \bibinfo{author}{\bibfnamefont{Y.}~\bibnamefont{Kuwahara}}, \bibnamefont{and}
  \bibinfo{author}{\bibfnamefont{M.}~\bibnamefont{Aono}},
  \bibinfo{journal}{Phys. Rev. B} \textbf{\bibinfo{volume}{52}},
  \bibinfo{pages}{R8711} (\bibinfo{year}{1995}).

\bibitem[{\citenamefont{Kim et~al.}(2002)\citenamefont{Kim, Kim, Hwang,
  Shresta, and Park}}]{Kim2002}
\bibinfo{author}{\bibfnamefont{Y.~K.} \bibnamefont{Kim}},
  \bibinfo{author}{\bibfnamefont{J.~S.} \bibnamefont{Kim}},
  \bibinfo{author}{\bibfnamefont{C.~C.} \bibnamefont{Hwang}},
  \bibinfo{author}{\bibfnamefont{S.~P.} \bibnamefont{Shresta}},
  \bibnamefont{and} \bibinfo{author}{\bibfnamefont{C.~Y.} \bibnamefont{Park}},
  \bibinfo{journal}{Surf. Sci.} \textbf{\bibinfo{volume}{498}},
  \bibinfo{pages}{116} (\bibinfo{year}{2002}).

\bibitem[{\citenamefont{Kinoshita et~al.}(1987)\citenamefont{Kinoshita, Kono,
  and Nagayoshi}}]{Kinoshita1987}
\bibinfo{author}{\bibfnamefont{T.}~\bibnamefont{Kinoshita}},
  \bibinfo{author}{\bibfnamefont{S.}~\bibnamefont{Kono}}, \bibnamefont{and}
  \bibinfo{author}{\bibfnamefont{H.}~\bibnamefont{Nagayoshi}},
  \bibinfo{journal}{J. of the Phys. Soc. of Japan}
  \textbf{\bibinfo{volume}{56}}, \bibinfo{pages}{2511} (\bibinfo{year}{1987}).

\bibitem[{\citenamefont{Kim et~al.}(2001)\citenamefont{Kim, Kim, Hwang,
  Shresta, An, and Park}}]{Kim2001}
\bibinfo{author}{\bibfnamefont{Y.~K.} \bibnamefont{Kim}},
  \bibinfo{author}{\bibfnamefont{J.~S.} \bibnamefont{Kim}},
  \bibinfo{author}{\bibfnamefont{C.~C.} \bibnamefont{Hwang}},
  \bibinfo{author}{\bibfnamefont{S.~P.} \bibnamefont{Shresta}},
  \bibinfo{author}{\bibfnamefont{K.~S.} \bibnamefont{An}}, \bibnamefont{and}
  \bibinfo{author}{\bibfnamefont{C.~Y.} \bibnamefont{Park}},
  \bibinfo{journal}{J. of the Korean Phys. Soc.} \textbf{\bibinfo{volume}{39}},
  \bibinfo{pages}{1032} (\bibinfo{year}{2001}).

\bibitem[{\citenamefont{Didiot et~al.}(2006)\citenamefont{Didiot,
  Fagot-Revurat, Pons, Kierren, Chatelain, and Malterre}}]{Didiot2006}
\bibinfo{author}{\bibfnamefont{C.}~\bibnamefont{Didiot}},
  \bibinfo{author}{\bibfnamefont{Y.}~\bibnamefont{Fagot-Revurat}},
  \bibinfo{author}{\bibfnamefont{S.}~\bibnamefont{Pons}},
  \bibinfo{author}{\bibfnamefont{B.}~\bibnamefont{Kierren}},
  \bibinfo{author}{\bibfnamefont{C.}~\bibnamefont{Chatelain}},
  \bibnamefont{and} \bibinfo{author}{\bibfnamefont{D.}~\bibnamefont{Malterre}},
  \bibinfo{journal}{Phys. Rev. B} \textbf{\bibinfo{volume}{74}},
  \bibinfo{pages}{081404} (\bibinfo{year}{2006}).

\bibitem[{\citenamefont{Moreschini et~al.}(2009)\citenamefont{Moreschini,
  Bendounan, Gierz, Ast, Mirhosseini, H\"{o}chst, Kern, Henk, Ernst, Ostanin
  et~al.}}]{MoreschiniSb2009}
\bibinfo{author}{\bibfnamefont{L.}~\bibnamefont{Moreschini}},
  \bibinfo{author}{\bibfnamefont{A.}~\bibnamefont{Bendounan}},
  \bibinfo{author}{\bibfnamefont{I.}~\bibnamefont{Gierz}},
  \bibinfo{author}{\bibfnamefont{C.~R.} \bibnamefont{Ast}},
  \bibinfo{author}{\bibfnamefont{H.}~\bibnamefont{Mirhosseini}},
  \bibinfo{author}{\bibfnamefont{H.}~\bibnamefont{H\"{o}chst}},
  \bibinfo{author}{\bibfnamefont{K.}~\bibnamefont{Kern}},
  \bibinfo{author}{\bibfnamefont{J.}~\bibnamefont{Henk}},
  \bibinfo{author}{\bibfnamefont{A.}~\bibnamefont{Ernst}},
  \bibinfo{author}{\bibfnamefont{S.}~\bibnamefont{Ostanin}},
  \bibnamefont{et~al.}, \bibinfo{journal}{Phys. Rev. B}
  \textbf{\bibinfo{volume}{79}}, \bibinfo{pages}{075424}
  (\bibinfo{year}{2009}).

\bibitem[{\citenamefont{Fu}(2009)}]{Fu2009}
\bibinfo{author}{\bibfnamefont{L.}~\bibnamefont{Fu}}, \bibinfo{journal}{Phys.
  Rev. Lett.} \textbf{\bibinfo{volume}{103}}, \bibinfo{pages}{266801}
  (\bibinfo{year}{2009}).

\bibitem[{NFE()}]{NFE}
\bibinfo{note}{According to Premper et al. \cite{Premper2007}, the orientation
  of the finite out-of-plane component of $\mathbf{P}$ breaks the sixfold
  symmetry of the underlying contours, by removing the mirror planes which
  point along the tips of the outer blossom-like CE contours. These generic
  results can be applied to our system which has the same symmetry. The
  orientation of the Bi trimers leaves $\overline{\Gamma\textmd{KM}}$ as the
  only mirror plane. As a result, the symmetry-required orientation of
  $\mathbf{P}$ is satisfied only if the outer hexagon tips lie within the
  $\overline{\Gamma\textmd{M}\Gamma}$ rather than the
  $\overline{\Gamma\textmd{KM}}$ directions as in Fig. 7 (c).}

\bibitem[{\citenamefont{Wan et~al.}(1992)\citenamefont{Wan, Guo, Ford, and
  Hermanson}}]{Wan1992}
\bibinfo{author}{\bibfnamefont{K.~J.} \bibnamefont{Wan}},
  \bibinfo{author}{\bibfnamefont{T.}~\bibnamefont{Guo}},
  \bibinfo{author}{\bibfnamefont{W.~K.} \bibnamefont{Ford}}, \bibnamefont{and}
  \bibinfo{author}{\bibfnamefont{J.~C.} \bibnamefont{Hermanson}},
  \bibinfo{journal}{Surf. Sci.} \textbf{\bibinfo{volume}{261}},
  \bibinfo{pages}{69} (\bibinfo{year}{1992}).

\bibitem[{\citenamefont{Kim et~al.}(2009)\citenamefont{Kim, Kim, McChesney,
  Rotenberg, Hwang, Hwang, and Yeom}}]{Kim2009}
\bibinfo{author}{\bibfnamefont{J.~K.} \bibnamefont{Kim}},
  \bibinfo{author}{\bibfnamefont{K.~S.} \bibnamefont{Kim}},
  \bibinfo{author}{\bibfnamefont{J.~L.} \bibnamefont{McChesney}},
  \bibinfo{author}{\bibfnamefont{E.}~\bibnamefont{Rotenberg}},
  \bibinfo{author}{\bibfnamefont{H.~N.} \bibnamefont{Hwang}},
  \bibinfo{author}{\bibfnamefont{C.~C.} \bibnamefont{Hwang}}, \bibnamefont{and}
  \bibinfo{author}{\bibfnamefont{H.~W.} \bibnamefont{Yeom}},
  \bibinfo{journal}{Phys. Rev. B} \textbf{\bibinfo{volume}{80}},
  \bibinfo{pages}{075312} (\bibinfo{year}{2009}).

\bibitem[{\citenamefont{Lee and Kang}(2008)}]{Lee2008}
\bibinfo{author}{\bibfnamefont{J.~Y.} \bibnamefont{Lee}} \bibnamefont{and}
  \bibinfo{author}{\bibfnamefont{M.~H.} \bibnamefont{Kang}},
  \bibinfo{journal}{J. Korean Phys. Soc.} \textbf{\bibinfo{volume}{53}},
  \bibinfo{pages}{3671} (\bibinfo{year}{2008}).

\bibitem[{\citenamefont{Kane and Mele}(2005)}]{Kane2005}
\bibinfo{author}{\bibfnamefont{C.~L.} \bibnamefont{Kane}} \bibnamefont{and}
  \bibinfo{author}{\bibfnamefont{E.~J.} \bibnamefont{Mele}},
  \bibinfo{journal}{Phys. Rev. Lett.} \textbf{\bibinfo{volume}{95}},
  \bibinfo{pages}{146802} (\bibinfo{year}{2005}).

\bibitem[{\citenamefont{Liu et~al.}(2009{\natexlab{a}})\citenamefont{Liu,
  Zhang, Wang, and Li}}]{Liu2009}
\bibinfo{author}{\bibfnamefont{G.}~\bibnamefont{Liu}},
  \bibinfo{author}{\bibfnamefont{P.}~\bibnamefont{Zhang}},
  \bibinfo{author}{\bibfnamefont{Z.}~\bibnamefont{Wang}}, \bibnamefont{and}
  \bibinfo{author}{\bibfnamefont{S.-S.} \bibnamefont{Li}},
  \bibinfo{journal}{Phys. Rev. B} \textbf{\bibinfo{volume}{79}},
  \bibinfo{pages}{035323} (\bibinfo{year}{2009}{\natexlab{a}}).

\bibitem[{\citenamefont{Liu et~al.}(2009{\natexlab{b}})\citenamefont{Liu, Wang,
  and Li}}]{Liu22009}
\bibinfo{author}{\bibfnamefont{G.}~\bibnamefont{Liu}},
  \bibinfo{author}{\bibfnamefont{Z.}~\bibnamefont{Wang}}, \bibnamefont{and}
  \bibinfo{author}{\bibfnamefont{S.-S.} \bibnamefont{Li}},
  \bibinfo{journal}{Physics Letters A} \textbf{\bibinfo{volume}{373}},
  \bibinfo{pages}{2091} (\bibinfo{year}{2009}{\natexlab{b}}).

\bibitem[{\citenamefont{R\"{u}egg et~al.}(2010)\citenamefont{R\"{u}egg, Wen,
  and Fiete}}]{Ruegg2009}
\bibinfo{author}{\bibfnamefont{A.}~\bibnamefont{R\"{u}egg}},
  \bibinfo{author}{\bibfnamefont{J.}~\bibnamefont{Wen}}, \bibnamefont{and}
  \bibinfo{author}{\bibfnamefont{G.~A.} \bibnamefont{Fiete}},
  \bibinfo{journal}{Phys. Rev. B} \textbf{\bibinfo{volume}{81}},
  \bibinfo{pages}{205115} (\bibinfo{year}{2010}).

\bibitem[{TBA()}]{TBAu}
\bibinfo{note}{Notice that the TB model (Eq. (10)) with the same power-law
  expression for the SO parameters, would yield a momentum splitting of only
  0.012 \AA$^{-1}$ for the Au(111) surface state \cite{FrantzeskakisThesis}, in
  agreement with the experimental results \cite{Reinert2001}. This proves that
  the giant splitting of the Bi/Si(111) system cannot be due to the magnitude
  of $\lambda_{ij}$.}

\bibitem[{\citenamefont{Oguchi and Shishidou}(2009)}]{Oguchi2009}
\bibinfo{author}{\bibfnamefont{T.}~\bibnamefont{Oguchi}} \bibnamefont{and}
  \bibinfo{author}{\bibfnamefont{T.}~\bibnamefont{Shishidou}},
  \bibinfo{journal}{J. Phys.: Condens. Matter} \textbf{\bibinfo{volume}{21}},
  \bibinfo{pages}{092001} (\bibinfo{year}{2009}).

\bibitem[{\citenamefont{Hsieh et~al.}(2009)\citenamefont{Hsieh, Xia, Wray,
  Qian, Pal, Dil, Osterwalder, Meier, Bihlmayer, Kane et~al.}}]{Hsieh2009}
\bibinfo{author}{\bibfnamefont{D.}~\bibnamefont{Hsieh}},
  \bibinfo{author}{\bibfnamefont{Y.}~\bibnamefont{Xia}},
  \bibinfo{author}{\bibfnamefont{L.}~\bibnamefont{Wray}},
  \bibinfo{author}{\bibfnamefont{D.}~\bibnamefont{Qian}},
  \bibinfo{author}{\bibfnamefont{A.}~\bibnamefont{Pal}},
  \bibinfo{author}{\bibfnamefont{J.~H.} \bibnamefont{Dil}},
  \bibinfo{author}{\bibfnamefont{J.}~\bibnamefont{Osterwalder}},
  \bibinfo{author}{\bibfnamefont{F.}~\bibnamefont{Meier}},
  \bibinfo{author}{\bibfnamefont{G.}~\bibnamefont{Bihlmayer}},
  \bibinfo{author}{\bibfnamefont{C.~L.} \bibnamefont{Kane}},
  \bibnamefont{et~al.}, \bibinfo{journal}{Science}
  \textbf{\bibinfo{volume}{323}}, \bibinfo{pages}{919} (\bibinfo{year}{2009}).

\bibitem[{\citenamefont{Frantzeskakis}(2010)}]{FrantzeskakisThesis}
\bibinfo{author}{\bibfnamefont{E.}~\bibnamefont{Frantzeskakis}}, Ph.D. thesis,
  \bibinfo{school}{EPFL Lausanne} (\bibinfo{year}{2010}).

\bibitem[{\citenamefont{Reinert et~al.}(2001)\citenamefont{Reinert, Nicolay,
  Schmidt, Ehm, and H\"{u}fner}}]{Reinert2001}
\bibinfo{author}{\bibfnamefont{F.}~\bibnamefont{Reinert}},
  \bibinfo{author}{\bibfnamefont{G.}~\bibnamefont{Nicolay}},
  \bibinfo{author}{\bibfnamefont{S.}~\bibnamefont{Schmidt}},
  \bibinfo{author}{\bibfnamefont{D.}~\bibnamefont{Ehm}}, \bibnamefont{and}
  \bibinfo{author}{\bibfnamefont{S.}~\bibnamefont{H\"{u}fner}},
  \bibinfo{journal}{Phys. Rev. B} \textbf{\bibinfo{volume}{63}},
  \bibinfo{pages}{115415} (\bibinfo{year}{2001}).

\end{thebibliography}
\end{document}